  \documentclass[aps,twocolumn,prb,showpacs]{revtex4}

\usepackage{graphicx}
\usepackage{latexsym}

\newcommand{\hsp}{\hspace*{-2mm}}

\begin{document}

\title{Universal and non-universal tails of distribution functions \\ in
the directed polymer and Kardar-Parisi-Zhang problems}
\author{I.V. Kolokolov}
\affiliation{L.D. Landau Institute for Theoretical Physics, Kosygina 2,
Moscow 119334, Russia}
\author{S.E. Korshunov}
\affiliation{L.D. Landau Institute for Theoretical Physics, Kosygina 2,
Moscow 119334, Russia}
\date{July 24, 2008}

\begin{abstract}

The optimal-fluctuation approach is applied to study the most distant
(non-universal) tails of the free-energy distribution function $P_L(F)$
for an elastic string (of a large but finite length $L$) interacting with
a quenched random potential. A further modification of this approach is
proposed which takes into account the renormalization effects and allows
one to study the closest (universal) parts of the tails. The problem is
analyzed for different dimensions of a space in which the polymer is
imbedded. In terms of the stochastic growth problem, the same distribution
function describes the distribution of heights in the regime of a
non-stationary growth in the situation when an interface starts to grow
from a flat configuration.

\end{abstract}

\maketitle

\section{Introduction}

A large variety of physical systems can be described in terms of an
elastic string interacting with a quenched random potential. The role of
such a string can be played by a domain wall in a two-dimensional magnet,
a vortex line in a superconductor, a dislocation in a crystal, and so on;
however following Ref. \onlinecite{KZ} the systems of such a kind are
usually discussed under the generic name of a directed polymer in a random
medium. The unfading interest to this problem is additionally supported by
its resemblance to more complex systems with quenched disorder (e.g., spin
glasses), as well as by its close relation to the dynamics of a randomly
stirred fluid and to the problem of a stochastic growth (see Refs.
\onlinecite{Kardar-R} and \onlinecite{HHZ} for reviews).

One of the main objects of interest in the directed polymer problem is
$P_L(F)$, the free-energy distribution function for large polymer length
$L$. In particular, the knowledge of this distribution function allows one
to make conclusions on the distribution of displacements. The first
important step in the analysis of $P_L(F)$ was made twenty years ago by
Kardar, \cite{Kardar} who  suggested that all moments of $P_L(F)$ can be
found by calculating the moments $Z_n\equiv \overline{Z^n}$ of the
distribution of the partition function $Z$ and proposed an asymptotically
exact method for the calculation of $Z_n$ in a $(1+1)\,$-dimensional
system (a string confined to a plane) with a $\delta$-correlated random
potential. However, soon after that Medina and Kardar \cite{MK} understood
(see also Ref. \onlinecite{DIGKB}) that the information provided by the
approach introduced in Ref. \onlinecite{Kardar} is insufficient for
finding any of the moments of $P_L(F)$. However, it allows one to find
\cite{Zhang} the tail of $P_L(F)$ at large negative $F$ (the left tail).
In such a situation the conclusions on the width of the distribution
function have to rely on the assumption that at large $L$ it acquires a
universal form,
\begin{equation}                                     \label{P*}
 P_L(F)=\frac{P_*(F/F_*)}{F_*}\;,
\end{equation}
incorporating the dependence on all parameters 
through a single characteristic free-energy scale $F_*(L)\propto
L^\omega$, which therefore can be extracted from the known form of the
tail. The form of Eq. (\ref{P*}) assumes that $F$, the free energy of a
directed polymer in a given realization of a disorder, {is counted off
from its average, $\bar{F}(L)$ or, more precisely, from the linear in $L$
contribution to $\bar F$ (that is, $L\lim_{L\rightarrow\infty}[\bar
F(L)/L]$). The same convention is implied below.}

Only recently it has been understood \cite{KK} that the form of the tail
following from Zhang's analysis \cite{Zhang} is applicable for the
description only of the most distant part of the left tail and therefore
has no direct relation to the universal form of the distribution function
which is achieved in the limit of $L\rightarrow\infty$. At large but
finite $L$ the form of $P_L(F)$ given by Eq. (\ref{P*}) can be expected to
be achieved only for not too large fluctuations of $F$ [that is, for
$|F|\ll F_c(L)$ with $F_c(L)/F_*(L)$ tending to infinity with the increase
of $L$], whereas the behavior of $P_L(F)$ at $|F|\gg F_c(L)$ remains
nonuniversal and is not obliged to have anything in common with $P_*(F/F_*)$.
In particular, it can incorporate quite different characteristic
free-energy scales. Thus, the fact that Zhang's approach \cite{Zhang}
reproduces both the correct form of the left tail of $P_*(F/F_*)$ and the
correct estimate of the universal free energy scale $F_*(L)$ (which is the
only relevant free-energy scale inside the universal region), is not more
than a happy coincidence. In contrast to that, the behavior of the right
tail of $P_L(F)$ inside the universal region is qualitatively different
from its behavior in the nonuniversal part of the tail. \cite{KK}

In this article the analysis of the universal and non-universal tails of
$P_L(F)$ developed in Ref. \onlinecite{KK} is presented in more detail and
also is extended to the investigation of $(1+d)\,$-dimensional systems, in
which polymer's displacement can be treated as a $d$-dimensional vector.
The article is organized as follows.

In Sec. \ref{II} we formulate the continuous model which is traditionally
applied for the description of the directed polymer problem and briefly
review its relation to the Kardar-Parisi-Zhang (KPZ) model \cite{KPZ}
of a stochastic growth, as well as to the Burgers turbulence problem.

Sec. \ref{OF} provides a short introduction to the optimal fluctuation
approach, which can be used for the description of the most distant
(non-universal) parts of the tails of $P_L(F)$. In Refs. \onlinecite{BFKL}
an analogous approach has been used to investigate the distribution of
velocity and its derivatives in the Burgers turbulence problem, which
however requires one to consider optimal fluctuations with completely
different structures then studied here.

In Sec. \ref{fl} the optimal fluctuation approach is applied for the
analysis of the far-left tail of $P_L(F)$, and in Sec. \ref{fr} of the
far-right tail. Our main attention is focused on the systems with a
$\delta$-correlated random potential; however for $d\geq 2$ the problem
with purely $\delta$-functional correlations becomes ill-defined, so we also
consider the case when a random potential correlations can be
characterized by a finite correlation radius.

For finding the universal parts of both tails one also has to look for
optimal fluctuations, but taking into account that in this regime the
parameters of the system have to be considered as scale dependent due to
their renormalization by fluctuations. This is done in Sec. \ref{ut}. The
validity of this approach is confirmed by the consistency of its
predictions with the results of the exact solution \cite{PS} of the
$(1+1)\,$-dimensional polynuclear growth (PNG) model, as well as by
obtaining identical estimates for $F_*(L)$ in the left and right tails.

The concluding Sec. \ref{Concl} is devoted to summarizing the results and
comparing them with some results of other authors, whereas in Appendix
\ref{RA} we discuss how some of the results of this work can be derived in
terms of the Kardar-Zhang replica approach. \cite{Kardar,Zhang}

Our main attention throughout this work is focused on a system with free
initial condition, that is, we assume that only one end of a string is
fixed, whereas the other one is free to fluctuate. In terms of the KPZ
problem \cite{KPZ} the same distribution function describes the
distribution of heights in the regime of a nonstationary growth in the
situation when an interface starts to grow from a flat configuration ($L$
being the total time of the growth). One only has to bear in mind that the
height (as defined in the standard form of the KPZ equation) and the free
energy of the directed polymer problem differ from each other by the sign.
Therefore, what we call here the left (right) tail of $P_L(F)$ in terms
of the KPZ problem corresponds to the right (left) tail of the height
distribution function.

Finally, Appendix \ref{FIC} is devoted to demonstrating that when
both end points of a directed polymer are fixed, the form of the left tail
of $P_L(F)$ remains basically the same as for free initial condition.

\section{The model \label{II}}

We consider an elastic string in a $(1+d)\,$-dimensional space
interacting with a random potential $V(t,{\bf x})$.
The coordinate along the average direction of the string is denoted $t$
for the reasons which will become evident few lines below.
Such a string can be described by the Hamiltonian,
\begin{equation}                                          \label{H}
H 
=\int_{0}^{t}dt' \left\{\frac{J}{2}\left[\frac{d{\bf x}(t')}{dt'}\right]^2
+V[t',{\bf x}(t')]\right\} \;,
\end{equation}
where the first term describes the elastic energy and the second one
the interaction with a random potential. Note that the form of the first
term in Eq. (\ref{H}) relies on the smallness of the angle between
the string and its preferred direction.

The partition function of a string which starts at $t=0$ and ends
at the point $(t,{\bf x})$ is then given by the functional integral,
\begin{equation}                                         \label{z(t,x)}
z(t,{\bf x})=\int_{-\infty}^{+\infty} d{\bf x}' \,z(0,{\bf x}')
\int_{{\bf x}(0)={\bf x}'}^{{\bf x}(t)={\bf x}}{\cal
D}{\bf x}(t')\exp\left(-H/T\right)\,,
\end{equation}
where $T$ is the temperature.
Naturally, $z(t,{\bf x})$ depends on the initial condition at $t=0$.
The fixed initial condition, ${\bf x}(t=0)={\bf x}_0$, corresponds to
$z(0,{\bf x})=\delta({\bf x}-{\bf x}_0)$,
whereas the free initial condition (which implies the absence of any
restrictions on ${\bf x}$ at $t=0$)  to
\begin{equation}                                            \label{FBC}
z(0,{\bf x})=\mbox{const}\,.
\end{equation}

Since Eq. (\ref{z(t,x)}) has exactly the same form as the Euclidean
functional integral describing the motion of a quantum particle whose mass
is given by $J$ in a time-dependent random potential $V(t,{\bf x})$ (with
$t$ playing the role of imaginary time and $T$ - of Plank's constant
$\hbar$), the evolution of $z(t,{\bf x})$ with the increase in $t$ has to
be governed by the imaginary-time Schr\"{o}dinger equation
\begin{equation}                                        \label{dz/dt}
-T\frac{\partial z}{\partial t}
=\left[-\frac{T^2}{2J}\nabla^2+V(t,{\bf x})\right]z(t,{\bf x})\,.
\end{equation}
As a consequence of this, the evolution of
the free energy corresponding to $z(t,{\bf x})$,
\begin{equation}                                        \label{}
f(t,{\bf x})=-T\ln\left[z(t,{\bf x})\right]\,,
\end{equation}
is governed \cite{HHF} by
the Kardar-Parisi-Zhang (KPZ) equation, \cite{KPZ}
\begin{equation}                                  \label{KPZ}
\frac{\partial f}{\partial t}+\frac{1}{2J}(\nabla f)^2-\nu \nabla^2 f
=V(t,{\bf x})\,,
\end{equation}
with the inverted sign of $f$, where $t$ plays the role of time
and $\nu\equiv{T}/{2J}$ of viscosity.
On the other hand, the derivation of Eq. (\ref{KPZ}) with respect
to ${\bf x}$ allows one to establish the equivalence \cite{HHF} between
the directed polymer problem and the Burgers equation \cite{Burgers}
with random potential force,
\begin{equation}                                 \label{Burg}
\frac{\partial {\bf u}}{\partial t}+\frac{1}{2}\nabla {\bf u}^2
-\nu\nabla^2{\bf u}=\frac{1}{J}\nabla V(t,{\bf x})\,,
\end{equation}
where the vector
\begin{equation}                                 \label{}
{\bf u}(t,{\bf x})\equiv\frac{1}{J}\nabla f(t,{\bf x})
\end{equation}
plays the role of velocity.
Note that in terms of the KPZ problem
the free initial condition (\ref{FBC})
corresponds to starting the growth from a flat
interface, $f(0,{\bf x})=\mbox{const}$, and in terms of the Burgers problem
to starting the evolution from a liquid at rest, ${\bf u}(0,{\bf x})=0$.

To simplify an analytical treatment, the statistic of a random potential 
is usually assumed to be Gaussian with
\begin{equation}                                              \label{VV}
\overline{V(t,{\bf x})}=0\,,~~~
\overline{V(t,{\bf x})V(t',{\bf x}')}=\delta(t-t')U({\bf x}-{\bf x}')\,,
\end{equation}
where an overbar denotes the average with respect to disorder.
{Our main attention below is focused on the case of purely
$\delta$-functional correlations, \makebox{$U({\bf x})=U_0\delta({\bf
x})$}. However, for $d\geq 2$ the problem with such a form of correlations
is ill-defined and needs a regularization, so we also consider the case when
$U({\bf x})$ can be characterized by a finite correlation radius $\xi$.
On the other hand, we always assume that the correlations in the $t$ direction
are $\delta$-functional, because in almost all situations considered below
the finiteness of the correlation radius in the $t$ direction can be ignored
as soon as it is small in comparison with the total length of
a string.}

\section{Optimal-fluctuation approach\label{OF}}

When the distribution of $V(t,{\bf x})$ is Gaussian and satisfies Eqs.
(\ref{VV}), the probability of any realization of $V(t,{\bf x})$ is
proportional to $\exp[-S\{V\}]$, where the action $S\{V\}$ is given by the
functional
\begin{equation}                                         \label{S(V)}
S\{V\}
= \frac{1}{2}\int_{0}^{L} dt
\int\hsp\int d{\bf x}\,d{\bf x}'\,
V(t,{\bf x})U^{-1}({\bf x}-{\bf x}')V(t,{\bf x}')\,.
\end{equation}
Here $U^{-1}({\bf x})$ denotes the function whose
convolution with $U({\bf x})$ is equal to $\delta({\bf x})$.
Accordingly, the probability of any time evolution of $f(t,{\bf x})$ is
determined by the action $S\{f\}$,
which is obtained by replacing $V(t,{\bf x})$ in Eq. (\ref{S(V)})
by the left-hand side of the KPZ equation (\ref{KPZ}).

To find the most optimal fluctuation having the largest probability
(in literature  it is often called ``instanton"),
one has to minimize $S\{f\}$ for the given boundary conditions at $t=0$
and $t=L$.
A convenient way to perform such a minimization consists in replacing
$S\{f\}$  by
\begin{eqnarray}                                         \label{S(f,mu)}
S\{f,\mu\} \hsp & = &\hsp \int_{0}^{L} dt\left\{\int d{\bf x}\,
\left[\frac{\partial f}{\partial t}+\frac{1}{2J}(\nabla f)^2
-\nu \nabla^2 f\right]\mu(t,{\bf x})\right. \nonumber\\
 \hsp &-&\hsp \left.\frac{1}{2}\int\hsp\int d{\bf x}\,d{\bf x}'\,
\mu(t,{\bf x})U({\bf x}-{\bf x}')\mu(t,{\bf x}')\right\}
\end{eqnarray}
where $\mu(t,{\bf x})$ is an auxiliary field with respect to which
$S\{f,\mu\}$ also has to be extremized.
Variation of Eq. (\ref{S(f,mu)}) with respect to $\mu(t,{\bf x})$
reproduces the KPZ equation (\ref{KPZ}) with
\begin{equation}                                       \label{V(mu)}
V(t,{\bf x})=\int d{\bf x}'\,U({\bf x}-{\bf x}')\mu(t,{\bf x}')\,,
\end{equation}
whereas its variation with respect to $f(t,{\bf x})$ leads to
\begin{equation}                                        \label{dmu/dt}
  {\partial \mu}/{\partial t}+\mbox{div}({\bf u}\mu)+\nu\nabla^2\mu = 0\,,                       \label{mu_t}
\end{equation}
where ${\bf u}(t,{\bf x})\equiv \nabla f(t,{\bf x})/J$ is the ``velocity''
entering the Burgers equation (\ref{Burg}).
The form of  Eq. (\ref{mu_t}) implies that the integral of $\mu(t,{\bf x})$
over whole space is a conserved quantity, whereas substitution
of Eq. (\ref{V(mu)}) into Eq. (\ref{S(V)})
shows that in terms of $\mu(t,{\bf x})$ the action can be rewritten as
\begin{equation}                                   \label{S(mu)}
S\{\mu\}=\frac{1}{2}\int_{0}^{L} dt \int
\int d{\bf x}\,d{\bf x}'\,\mu(t,{\bf x})U({\bf x}-{\bf x}')\mu(t,{\bf x}')\,.
\end{equation}
In a system with $\delta$-functional correlations,
\makebox{$U(x)=U_0\delta({\bf x})$}, $V$ and $\mu$ differ from each other
only by a constant factor $U_0$, and accordingly,
Eq. (\ref{mu_t}) can be replaced by
\begin{equation}                                        \label{V_t}
    {\partial V}/{\partial t}+\mbox{div}({\bf u} V)+\nu \nabla^2 V=0\,.
\end{equation}

If the beginning of a polymer (at $t=0$)  is not fastened to a particular
point and is free to fluctuate, the initial condition for the partition
function $z(t,{\bf x})$ has to be chosen in the form
$z(0,{\bf x})=\mbox{const}$.
In such a case to find the tails of $P_L(F)$
one has to find the solution of Eqs. (\ref{KPZ}) and (\ref{mu_t})
which satisfies the initial condition
\begin{equation}                                        \label{ini}
f(0,{\bf x})=0\,,
\end{equation}
and the final condition
\begin{equation}                                        \label{fin}
f(L,0)=F\,,
\end{equation}
where for the left tail $F<0$ and for the right tail $F>0$. Alternatively,
condition (\ref{fin}) can be imposed by the inclusion of the
$\delta$-functional factor,
\begin{equation}                                        \label{}
\int d\lambda\exp\{ i\lambda[f(L,{\bf x}=0)-F]\}\,,
\end{equation}
into the functional integral defining the probability of a fluctuation.
In such a case condition (\ref{fin}) for $f(L,{\bf x})$
should be replaced by the condition for $\mu(L,{\bf x})$,
\begin{equation}                                        \label{fin-2}
\mu(L,{\bf x})=\mu_0\delta({\bf x})\,,
\end{equation}
where, however, the value of
$\mu_0\propto\lambda$ has to be chosen to satisfy Eq. (\ref{fin}).

\section{Far-left tail \label{fl}} 

It turns out that in the case of the left tail
the solution of Eqs. (\ref{KPZ}) and (\ref{mu_t})
which satisfies boundary conditions (\ref{ini}) and (\ref{fin})
can be constructed on the basis of the solution of these equations
in which the potential $V$ and all derivatives of $f$ do not depend on $t$,
which means that the time dependence of $f(t,{\bf x})$
is decoupled from its spacial dependence and is as trivial as possible,
\begin{equation}                                       \label{f(x)0}
f(t,{\bf x})={E}(t-t_1)+f({\bf x})\,,
\end{equation}
where $t_1=\mbox{const}$ and ${E}=\mbox{const}<0$.
Below we for brevity call such solutions stationary.

For $f(t,{\bf x})$ of form (\ref{f(x)0})
the replacement
\begin{equation}                                        \label{Psi(f)}
f({\bf x})=-T\ln\Psi({\bf x})\,
\end{equation}
transforms the KPZ equation (\ref{KPZ}) into a
stationary Schr\"{o}dinger equation:
\begin{equation}                                        \label{Schr}
{E}\Psi=\hat{H} 
\Psi\,,
\end{equation}
for a single-particle quantum-mechanical problem defined by the Hamiltonian
\begin{equation}                                       \label{Hq}
\hat{H}=-\frac{T^2}{2J}\nabla^2+V({\bf x})\,,
\end{equation}
where $J$ plays the role of mass and $T$ of Plank's constant $\hbar$
[compare with (\ref{dz/dt})].
On the other hand, when both \makebox{${\bf u}=-(T/J)\nabla\Psi/\Psi$}
and $\mu$  do not depend on $t$,
Eq. (\ref{mu_t}) is automatically fulfilled as soon as
\begin{equation}                                       \label{mu(psi)}
\mu({\bf x})\propto \Psi^2({\bf x})\,,
\end{equation}
which implies
\begin{equation}                                       \label{V(mu)2}
V({\bf x})=
-\Lambda\int_{}^{}d{\bf x}'\,U({\bf x}-{\bf x}')\Psi^2({\bf x}')\,,
\end{equation}
where $\Lambda$ is an arbitrary constant.
Substitution of Eq. (\ref{V(mu)2}) into Eq. (\ref{Schr}) allows one
to replace them by a single nonlinear Schr\"{o}dinger equation,
\[
{E}\Psi=-\frac{T^2}{2J}\nabla^2\Psi-\Lambda\Psi({\bf x})
\int_{-\infty}^{+\infty}d{\bf x}'\,U({\bf x}-{\bf x}')\Psi^2({\bf x}')\,.
\]

Equation (\ref{V(mu)2}) has been derived by Halperin and Lax \cite{HL}
when looking for the optimal fluctuation of the potential $V({\bf x})$,
which for the given value of the ground state energy \makebox{$E<0$}
of the quantum-mechanical Hamiltonian (\ref{Hq}) minimizes the functional
\begin{equation}                                         \label{s(V)}
s\{V\}
= \frac{1}{2}\int_{}^{}\int_{}^{}d{\bf x}\,d{\bf x}'\,
V({\bf x})U^{-1}({\bf x}-{\bf x}')V({\bf x}')\,,
\end{equation}
determining the probability of $V({\bf x})$
(or, equivalently,
minimizes $E$ for the given value of $s\{V\}$).
Apparently, in terms of our problem $s\{V\}$ is related to the action
$S\{V\}$ defined by Eq. (\ref{S(V)}) as $S=Ls$.
In the case of $\delta$-functional correlations,
$U({\bf x})=U_0\delta({\bf x})$,
and $t$-independent potential $V({\bf x})$,
functional (\ref{S(V)}) is reduced to
\begin{equation}                                        \label{S(V)2}
S\{V\}
= \frac{L}{2U_0}
\int_{}^{}d{\bf x}\,V^2({\bf x})\,.
\end{equation}

\subsection{$\delta$-functional correlations, $\bf d=1$\label{d=1}}

In a $1+1$-dimensional system with a $\delta$-correlated random potential,
$U(x)=U_0\delta(x)$,
the localized solution of Eqs. (\ref{Schr}) and (\ref{V(mu)2}) (the soliton)
exists for any ${E}<0$ and can be found exactly, \cite{HL}
\begin{eqnarray}                                           \label{Psi}
\Psi(x) & = &
\left(\frac{-2E}{\Lambda U_0}\right)^{1/2}\frac{1}{\cosh(x/\Delta)}\;, \\
V(x) & = & \frac{2E}{\cosh^2({x}/{\Delta})}\;,
                                                             \label{V(x)}
\end{eqnarray}
where the length-scale
\begin{equation}                                          \label{}
\Delta=\frac{T}{(-2J{E})^{1/2}}
\end{equation}
can be called soliton width.
This allows one to conclude that the stationary solution of
Eqs. (\ref{KPZ}) and (\ref{V_t}) is given by Eq. (\ref{V(x)}) and
\begin{eqnarray}                                             \label{f(x)}
f(t,x) & = & E(t-t_1)+T\ln\left(2\cosh\frac{x}{\Delta}\right) \;,
\end{eqnarray}
which follows from the substitution of Eq. (\ref{Psi})
into Eqs. (\ref{f(x)0}) and (\ref{Psi(f)}). Note that in Eq. (\ref{f(x)})
the constant $t_1$ has been redefined in order to absorb $\Lambda$.

\begin{figure}[bt]
\begin{center}
\includegraphics[width=50mm]{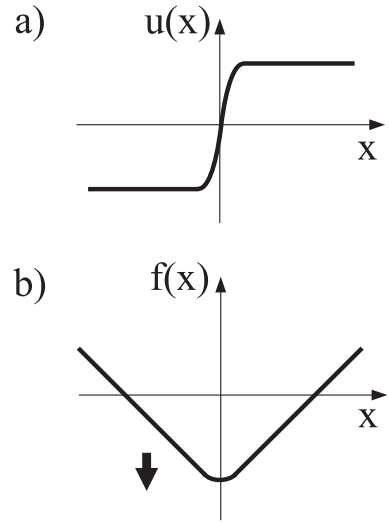}
\caption[Fig. 1] {The spacial dependence of $u$ and $f$
in the stationary  solution of Eqs. (\ref{KPZ})
and (\ref{V_t}).}\label{fig1}
\end{center}
\end{figure}

Differentiation of Eq. (\ref{f(x)}) with respect to $x$ gives
a stationary profile of $u(x)$,
\begin{equation}                                   \label{u(x)}
    u(x)=v\tanh\frac{x}{\Delta} \;,
\end{equation}
schematically shown in Fig. \ref{fig1}(a).
Here
\begin{equation}                                   \label{}
v={T}/{J\Delta}
\end{equation}
is the velocity of the outward flow created by the forces
acting inside the soliton. The profile (\ref{u(x)})
up to a sign coincides with the one
in a stationary shock wave with the same amplitude $v$.
The solitons of such a kind (both stationary and moving ) have been
discussed in a number of works by Fogedby.\cite{Fogedby}

The stationary profile of $f$ described by Eq. (\ref{f(x)})
is schematically shown in Fig. \ref{fig1}(b).
With the increase in time it is moving downward as a whole with a constant
velocity ${\partial f}/{\partial t}=E$.
Away from the soliton's core, that is at $|x|\gg \Delta$,
the dependence described by Eq. (\ref{f(x)}) can be approximated as
\begin{equation}                                      \label{f(x)-appr}
f(t,x) \approx E(t-t_1)+(-2JE)^{1/2}|x|\,.
\end{equation}
Since Eq. (\ref{f(x)-appr}) describes a solution of the noiseless KPZ
equation, its form does not depend on the form (or amplitude)
of the random potential correlator $U(x)$.

The stationary solution minimizes the action
for the given negative value of ${\partial f}/{\partial t}=E$.
Therefore, it allows one to find the optimal value of $S$ in situations
when it is not influenced by the initial condition.
Substitution of Eq. (\ref{V(x)}) into Eq. (\ref{S(V)2}) then gives
\begin{equation}                                      \label{}
S(\Delta)=\frac{2}{3}\frac{T^4L}{U_0J^2\Delta^3}\,.
\end{equation}
Apparently, the condition $f(L,0)-f(0,0)=F$ is fulfilled when
$E=F/L$, which corresponds to
\begin{equation}                                       \label{F(Delta)}
    F=-\frac{T^2}{2J\Delta^2}L
\end{equation}
and
\begin{equation}                                       \label{S(F)}
S(F)=\frac{4\sqrt{2}
}{3}\frac{\;T{(-F)^{3/2}}}{U_0J^{1/2}{L^{1/2}}}\;.
\end{equation}

However, the real optimal fluctuation also
has to respect the initial condition and it is clear that
the spacial dependence of $f$ in Eq. (\ref{f(x)})
in no way resembles the initial condition (\ref{ini}).
In terms of the quantum-mechanical problem with time-independent
potential $V({\bf x})$ it is clear that the applicability of the relation
$F\approx E L$ requires to have $(\delta E)L \gg T$,
where $\delta E$ is the energy gap separating the ground state
of the Hamiltonian (\ref{Hq}) from the first excited state.
Since in potential (\ref{V(x)}) there exists only one bound level with
a negative energy, \cite{LL-QM} whereas excited states can have any
non-negative energy, this condition is equivalent to $-F \gg T$.

\begin{figure}[bt]
\begin{center}
\includegraphics[width=50mm]{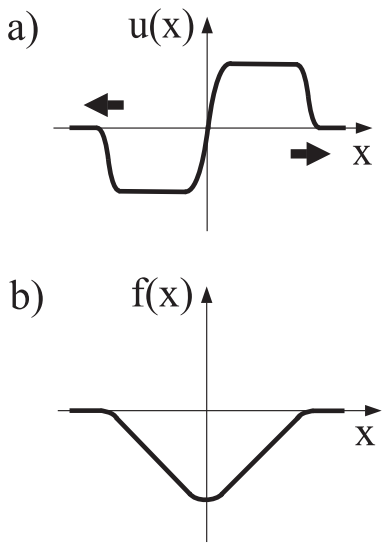}
\caption[Fig. 1] {The spacial dependence of $u$ and $f$ in the solution of
Eqs. (\ref{KPZ}) and (\ref{V_t}) corresponding to the left tail of
$P_L(F)$. The arrows show the directions of motion of the two shock waves.
}\label{fig2}
\end{center}
\end{figure}

For constructing a non-stationary solution which eliminates the inconsistency
between the forms of the stationary solution and of the initial condition
({without increasing the action}),
one has to complement the soliton shown
in Fig. \ref{fig1}(a) by two traveling shock waves
[as shown in Fig. \ref{fig2}(a)], whose existence  does not require
any additional pumping. Both these shock waves will be moving outwards
with velocity $v/2$. 
Their presence will change the profile of $f(t,x)$ to the one shown in Fig.
\ref{fig2}(b), so that  $f(x)$ will be given by Eq. (\ref{f(x)}) only
in the interval where $f(t,x)<0$, whereas outside of this region
it will coincide with the initial condition (\ref{ini}) (with a smooth
crossover between the two solutions).
This means that if a potential localized in the vicinity of
$x=0$ is switched on at $t=0$, its influence on $f(t,x)$ at $t>0$ extends
only to a finite (but growing with $t$) region,
which is perfectly logical.

In such a situation the constant $t_1$ in Eq. (\ref{f(x)}) [or in Eq.
(\ref{f(x)-appr})] will have the meaning of an effective time required for
the formation of the non-stationary solution shown in Fig. \ref{fig2}. At
the initial stage, that is, at $t\lesssim t_1$, the spacial distribution
of $V(x)$ will substantially differ from the one given by Eq. (\ref{V(x)})
The value of $t_1$ can be estimated from the comparison of the soliton
width $\Delta=2\nu/v$ with the velocity $v$ of the flow it creates, which
gives $t_1\sim\nu/v^2$. This allows one to expect that for $L\gg t_1$ the
main contribution to the action is coming from the region in $t$ where Eq.
(\ref{V(x)}) gives a sufficiently accurate description of the solution,
and therefore the value of $S(F)$ is given by Eq. (\ref{S(F)}). In terms
of $F$ the constraint $L\gg t_1$ corresponds to the condition $-F\gg T$
(which was already derived above in different terms).

The same condition allows one to neglect the final stage of the
optimal-fluctuation evolution. At this stage the potential has to shrink
from the form given by Eq. (\ref{V(x)}) to a $\delta$-function as suggested
by Eq. (\ref{fin-2}). Simultaneously, the downward tip of $f(x)$ has to
change its shape from rounded to more sharp.
The decrease in $f(x=0)$ related  to this process can be expected
to be comparable with the change in $f$ induced by the rounding of the tip,
which according to Eq. (\ref{f(x)}) is of the order of $T$, and therefore
for $-F\gg T$ can be ignored.

Note that the same answer for $S(F)$, Eq. (\ref{S(F)}), can be also obtained
in the framework of the Kardar-Zhang replica approach 
based on mapping a system to a set of interacting bosons and keeping only
the ground state contribution to the partition function of these bosons
(see Appendix \ref{RA} for more details).

In Appendix \ref{FIC} we demonstrate that the change of the initial
condition from free to fixed does not change the form
of the main contribution to $S(F)$. The same conclusion is even more
easily attained in terms of the replica approach (see Appendix
\ref{RA}).

In the remaining part of this section, we analyze the systems with
an arbitrary dimension and/or finite-range correlations assuming that
the main features of the optimal fluctuation determining the far-left tail
are the same.
Namely, we expect that in a growing region around \makebox{${\bf x}=0$},
the solution is close to the stationary solution, whereas outside of this
region it is close to the initial condition $f(t,{\bf x})=0$, the crossover
between the two regions being described by a corresponding solution of
the noiseless KPZ equation. In such a situation the action of the optimal
fluctuation is determined by the form of the stationary solution.

\subsection{Generalization to $\bf d\neq 1$\label{fl-d}}

If the dimension of the transverse space $d$ is not equal to 1,
the joint solution of Eqs. (\ref{Schr}) and (\ref{V(mu)2}),
that is, the wave function $\Psi({\bf x})$
which minimizes the sum of a positive kinetic energy,
\begin{equation}                                             \label{}
{\cal K}\equiv\frac{T^2}{2J}
\frac{\int_{}^{}d^d{\bf x}|\nabla\Psi({\bf x})|^2}
{\int_{}^{}d^d{\bf x}|\Psi({\bf x})|^2}\;,
\end{equation}
and a negative potential energy,
\begin{equation}                                            \label{}
{\cal V}\equiv \frac{\int_{}^{}d^d{\bf x}V({\bf x})|\Psi({\bf x})|^2}
{\int_{}^{}d^d{\bf x}|\Psi({\bf x})|^2}\;,
\end{equation}
for a given value of the functional $S\{V\}$ defined by Eq. (\ref{S(V)2}),
cannot be found exactly.
However, in a situation when this wave function 
and, therefore, the potential $V({\bf x})\propto -U_0\Psi^2({\bf x})$
are well localized at some length scale $\Delta$,
an estimate for $\Delta$ and a qualitative relation between $S$ and $F$
can be obtained without finding the exact form of $\Psi({\bf x})$.

When $\Psi({\bf x})$ 
can be characterized by a single relevant length-scale $\Delta$, one has
\begin{equation}                                       \label{cal-K}
{\cal K}(\Delta)\sim\frac{T^2}{J\Delta^2}\;,
\end{equation}
whereas 
the absolute value of \makebox{${\cal V}\sim V(0)$} at a given $S$
can be estimated with the help of Eq. (\ref{S(V)2}), which gives
\begin{equation}                                        \label{S(cal-V)}
S 
 \sim \frac{L}{U_0}\Delta^d {\cal V}^2\,,
\end{equation}
and therefore
\begin{equation}                                        \label{V(S)}
{\cal V}(\Delta)
\sim -\left(\frac{SU_0}{L\Delta^d}\right)^{1/2}\,.
\end{equation}
For $0<d<4$ the sum ${\cal K}(\Delta)+{\cal V}(\Delta)$ has a minimum
with respect to $\Delta$ when
\makebox{${\cal K}(\Delta)\sim -{\cal V}(\Delta)$} and therefore both
${\cal K}$ and $-{\cal V}$ have to be of the same order
as $-E=-F/L$.

Substitution of ${\cal V}\sim F/L$ into Eq. (\ref{S(cal-V)})
allows one to rewrite this relation as
\begin{equation}                              \label{S(Delta,F)}
S\sim \frac{\;\,\Delta^dF^2}{U_0L}\;.
\end{equation}
On the other hand, an estimate for $\Delta$ in terms of $F$
can be obtained from the relation ${\cal K}\sim -E$, which gives
\begin{equation}                                          \label{Delta(F)}
\Delta(F)\sim \frac{T}{2J}\left[\frac{JL}{-F}\right]^{1/2}\,.
\end{equation}
After that to obtain an estimate for $S(F)$
one needs only to substitute Eq. (\ref{Delta(F)}) into Eq.
(\ref{S(Delta,F)}), which leads to
\begin{equation}                                   \label{S(F)d}
S(F)\sim \frac{\;\;T^d{\left(-F\right)^{2-d/2}}}{U_0J^{d/2}{L^{1-d/2}}}\;.
\end{equation}
Naturally, for $d=1$ Eq. (\ref{S(F)d}) is consistent with Eq. (\ref{S(F)})
derived in the Sec. \ref{d=1} on the basis of the exact solution of
Eqs. (\ref{Schr}) and (\ref{V(mu)2}).

For $d>4$ the sum of ${\cal K}(\Delta)$ and ${\cal V}(\Delta)$ at a given $S$
is not bounded from below and tends to $-\infty$ when $\Delta\rightarrow 0$.
Accordingly, for any $F<0$ it becomes possible to find
a stationary fluctuation with an arbitrary low action,
so the method of optimal fluctuation is no longer applicable.
However, it turns out that the range of the applicability of Eq. (\ref{S(F)d})
is even more narrow than the interval $0<d<4$, where the action of
stationary fluctuations has a well-defined positive minimum.

The point is that $L$ enters Eq. (\ref{S(F)d}) as the total time
of the development of the optimal fluctuation of $f(t,{\bf x})$.
From this it is clear that Eq. (\ref{S(F)d}) can be expected to be valid
only if $S$ decreases with the increase in $L$,
which forces the time of the development of the optimal fluctuation
to coincide with $L$.
In the opposite case (when $S$ {decreases} with the {\em decrease} of $L$)
there appears a possibility to decrease the action of the fluctuation
we are considering by making the time of its development smaller than $L$.
Namely, if one makes in Eq. (\ref{S(Delta,F)}) a replacement
\begin{equation}                                                 \label{}
L\Rightarrow\gamma^2L\,,~~~
\Delta\Rightarrow\gamma\Delta
\end{equation}
conserving relation (\ref{Delta(F)}), this leads to 
$S\Rightarrow\gamma^{d-2}S$. Therefore, for $d>2$ a consistent decrease
in the size of the fluctuation and in the time of its development allow
one to make $S(F)$ {arbitrarily small}
by choosing a sufficiently small $\gamma$.
This suggests that the result (\ref{S(F)d}) can be expected
to be applicable only at $0<d<2$, whereas at $d>2$
the optimal fluctuation corresponding to the most
distant part of the left tail has to be localized at small scales and its
form has to be determined by the form of a cutoff.
Without a cutoff the problem with $d\geq 2$ and $\delta$-functional
correlations is ill-defined.

Note that at $d\geq 2$, the problem with $\delta$-functional
correlations is ill-defined also for another reason.
Namely, at $d\geq 2$  the perturbative corrections to the viscosity
$\nu$ and other quantities acquire ultraviolet divergencies which at $d<2$
are absent.
Apparently, this is not a coincidence  but another manifestation
of the same phenomenon. Therefore, for $d\geq 2$ some ultraviolet cutoff
must be introduced into the problem.
One of the most natural ways to do it consist in assuming that
the correlations of a random potential are characterized by
a finite correlation radius.

\subsection{Finite-range correlations \label{frc}}

When random potential correlator $U({\bf x})$ [which we assume to be
spherically symmetric, $U({\bf x})\equiv U({|\bf x}|)$] is characterized
by a finite correlation radius $\xi$,
the stationary solution of Eqs. (\ref{KPZ}) and (\ref{mu_t})
cannot be found exactly even at $d=1$.
However, it is clear from the form of Eq. (\ref{V(mu)}) relating $V$ and
$\mu$ that when the soliton width $\Delta$ is much larger than
$\xi$, the actual solution has to be rather close to the solution
for $\xi=0$, the same being true also for the value of $S(F)$.
It follows from Eq. (\ref{Delta(F)}) that in terms of $F$
the condition $\Delta\gg\xi$ corresponds to
\begin{equation}                                                \label{}
-F\ll F_\xi\sim \frac{T^2L}{J\xi^2}\;.
\end{equation}

It turns out that for the opposite relation between the parameters,
$-F\gg F_\xi$,
the stationary solution of Eqs. (\ref{KPZ}) and (\ref{mu_t})
also can be found rather accurately.
As it is shown below, in such a case $\mu$ 
is localized in a region which is much narrower than $\xi$,
whereas both $f$ and $V$ change at the scales of the order of $\xi$.
In particular, it follows from Eq. (\ref{V(mu)}) that in such a situation
the spacial dependence of the potential $V({\bf x})$ just repeats that of
$U({\bf x})$,
\begin{equation}                                   \label{V(e)}
    V({\bf x})\approx -U({\bf x}){\varepsilon}\,,
\end{equation}
whereas the amplitude of $V({\bf x})$ is determined by
\begin{equation}                                           \label{}
   \varepsilon\equiv -\int_{}^{}d{\bf x}\,\mu({\bf x})\,,
\end{equation}
the overall strength of the negative potential source $\mu({\bf x})$.

For $-F\gg F_\xi$ the viscous term in Eq. (\ref{KPZ})
can be neglected, which immediately gives that in the
spherically symmetric stationary solution
\begin{equation}                                            \label{}
\partial f/\partial t \approx -U(0)\varepsilon
\end{equation}
and
\begin{equation}                                           \label{}
\left(\frac{\partial f}{\partial r}\right)^2
={2J[U(0)-U(r)]\varepsilon}\,,
\end{equation}
where $r=|{\bf x}|$, so that
\begin{equation}                                  \label{f(r)}
    f({\bf x})=f(0)+\sqrt{2J\varepsilon}\int_{0}^{|{\bf x}|}dr\,
    \sqrt{U(0)-U(r)}\;.
\end{equation}
In terms of the Schr\"{o}dinger equation (\ref{Schr}) the neglect of the
viscous term in the stationary KPZ equation corresponds to nothing else
but using the semiclassical approximation
for the calculation of the ground-state wave function.

Substitution of $\Psi({\bf x})=\exp[-f({\bf x})/T]$ with $f({\bf x})$
given by Eq. (\ref{f(r)})
into Eq. (\ref{mu(psi)}) demonstrates that at $|{\bf x}|\ll\xi$,
\begin{equation}                                          \label{mu(x)}
\mu({\bf x})\propto\exp\left[-\frac{{\bf x}^2}{2\Delta^2}\right]\,,
\end{equation}
where  $\Delta$, the width of the region where the
potential source $\mu({\bf x})$ is localized, is given by
\begin{equation}                                   \label{x_mu}
\Delta(F)=
\left[\frac{T^2U(0)}{-4U_{rr}(0)JE}\right]^{1/4}
\sim \left(\frac{F_\xi}{-F}\right)^{1/4}\xi\,.
\end{equation}
When deriving this estimate we have replaced
$-U_{rr}(0)$  by $U(0)/\xi^2$ and $E$ by $F/L$.
The result shows that the assumption $\Delta(F)\ll\xi$,
which has been used above to obtain Eq. (\ref{V(e)}),
is indeed self-consistent as soon as \makebox{$-F\gg F_\xi$}.

Substitution of Eq. (\ref{V(e)}) into Eq. (\ref{S(mu)}) reduces
the expression for the action to a very simple form,
\begin{equation}                                         \label{S(e)}
S=\frac{U(0)}{2}{L}\varepsilon^2
 =\frac{LE^2}{2U(0)} \;,
\end{equation}
which is easily recognizable to those familiar with application of
the optimal-fluctuation approach to a quantum-mechanical problem 
with finite-range correlations of a random potential \cite{ShEf}
and after substitution of $E=F/L$ gives
\begin{equation}                                           \label{S(F)xi}
S(F)=\frac{F^2}{2U(0)L}\;.
\end{equation}
The same temperature-independent answer can be also reproduced in terms of
the Kardar-Zhang replica approach (see Appendix \ref{RA}).

Thus we have demonstrated that for $\xi>0$ the most distant part of the
left tail is Gaussian independently of the dimension. Since the width of
the region where $\mu$ is localized grows with the decrease in $-F$, a
crossover to some other regime must occur when this width becomes
comparable with $\xi$. In particular, for $d<2$ and \makebox{$\xi\ll x_0$}
the dependence of $S$ on $F$ at $-F\ll F_\xi$ has to be described by Eq.
(\ref{S(F)d}) with a subsequent crossover to the universal regime
discussed in Sec. \ref{ul}. Naturally, the increase in $\xi$ (or in $d$)
leads to shrinking and subsequent vanishing of the region where $S(F)$ can
be described by Eq. (\ref{S(F)d}). On the other hand, when $\xi$ is taken
to zero $F_\xi$ goes to infinity, which leads to the disappearance of the
region with Gaussian behavior.

\subsection{A boundary from below}

It is worthwhile to emphasize that expression (\ref{S(F)xi}) gives an exact
boundary from below for the value of $S(F)$ in the optimal fluctuation.
This is so because the potential of the form
\begin{equation}                                           \label{V(V0)}
V({\bf x})=\frac{U({\bf x})}{U(0)}{V({\bf x}=0)}
\end{equation}
minimizes functional (\ref{s(V)}) for the given value of $V({\bf x}=0)$,
from where
\begin{equation}                                           \label{}
S(F)\geq \frac{1}{2U(0)}\int_{0}^{L}dt\,[V(t,{0})]^2\,.
\end{equation}
On the other hand, in a growing fluctuation of $f(t,{\bf x})$
which has a spherically symmetric shape and an extremum at ${\bf x}=0$,
the absolute value of ${\partial f(t,{0})}/{\partial t}$ is bounded
from above by $|V(t,{0})|$ because at the point of extremum
the second term in the left-hand side of the KPZ equation (\ref{KPZ})
vanishes, whereas the third term, $-\nu \nabla^2 f$, has to have
the same sign as ${\partial f(t,{0})}/{\partial t}$.
This allows one to conclude that
\begin{eqnarray}                                          \nonumber
S(F) & \geq & \frac{1}{2U(0)}\int_{0}^{L}dt\, \left[\frac{\partial f(t,{0})}
{\partial t}\right]^2 \\
& \geq & \frac{F^2}{2U(0)L}                               \label{Smin}
\end{eqnarray}
Apparently, this inequality is reduced to equality only if
(i) $V(t,{\bf x})$ is of form (\ref{V(V0)}),
(ii) the viscous term in the KPZ equation can be neglected, and
(iii) ${\partial f(t,{0})}/{\partial t}$ does not depend on time.
Since in the negative fluctuation of $f$ considered in Sec. \ref{frc}
all these conditions are satisfied rather accurately,
the action of this fluctuation is approximately  equal to the boundary
from below given by Eq. (\ref{Smin}).

Note that the argument leading to the derivation of
Eq. (\ref{Smin}) is valid for {both signs} of $F$. Therefore
inequality (\ref{Smin}) has to be satisfied also in the far-right tail.

\section{Far-right tail \label{fr}} 

Our analysis has established that the optimal fluctuation corresponding to
the left tail of $P_L(F)$ has a very special shape which can be
characterized by two different scales. Namely, the size of the area where
the potential $V$ is localized, $\Delta(F)$, is much smaller then the total
size of the fluctuation
$\tilde{\Delta}(F)\sim\left({-FL}/{J}\right)^{1/2}$, that is,
the width of the area where $f$ and ${\bf u}$ essentially deviate from zero.
Apparently this property is closely related to the fact that inside a
growing negative fluctuation of $f$ the terms ${\partial f}/{\partial t}$
and $(1/2J)(\nabla f)^2$ in the functional,
\begin{equation}                                  \label{S(f)0}
S\{f\}=\frac{1}{2U_0}\int_{0}^{L}\hsp dt \int d{\bf x}\,
\left[\frac{\partial f}{\partial t}+\frac{1}{2J}(\nabla f)^2
      -\nu \nabla^2 f\right]^2\,
\end{equation}
defining the probability of a fluctuation in a system with a
$\delta$-correlated potential have to be of the opposite signs.
This provides a possibility for their mutual compensation
in almost the whole volume of the fluctuation.
It is clear that in the case of the right tail such a cancellation
is impossible because in the substantial part of the optimal fluctuation
${\partial f}/{\partial t}$ has to be of the same sign as
$(1/2J)(\nabla f)^2$.
As a consequence, the optimal fluctuation corresponding to the right tail
must have a shape which can be characterized
by a single relevant length-scale, $\Delta_+(F)$.

This length scale can be estimated from the comparison
of ${\partial f}/{\partial t}\sim F/L$ with
\makebox{$(1/2J)(\nabla f)^2\sim F^2/J\Delta_+^2$},
which shows that $\Delta_+$ has to be of the same order as the total size
of the optimal fluctuation with $F<0$:
\begin{equation}                                  \label{Delta(F)+}
\Delta_+(F)\sim \tilde{\Delta}(-F)\sim\left(\frac{LF}{J}\right)^{1/2}\!\!.
\end{equation}
Note that for $\Delta_+(F)$ given by Eq. (\ref{Delta(F)+})
the viscous term in the integrand of functional (\ref{S(f)0})
can be neglected if $F$ is large enough.
This is precisely the reason why an estimate for $\Delta_+$ can be obtained
by matching the two other terms in this integrand.
A comparison of $\nu\nabla^2f\sim \nu F/\Delta_+^2$
with $(1/2J)(\nabla f)^2\sim F^2/J\Delta_+^2$ shows that the condition
which allows one to neglect the viscous term can be written as
$F\gg 2J\nu=T$. Apparently this constraint is automatically fulfilled as
soon as one considers the most distant part of the tail.

Substitution of Eq. (\ref{Delta(F)+}) into the relation
\begin{equation}                                       \label{}
S\sim\frac{L\Delta_+^d}{U_0}\left(\frac{F}{L}\right)^2\,,
\end{equation}
following from the assumption that $\Delta_+(F)$ is the only relevant
length-scale in the problem,
gives then an estimate for the action determining the form of
the far-right tail of $P_L(F)$,
\begin{equation}                                \label{S(F)+}
S(F)\sim \frac{F^{2+d/2}}{U_0J^{d/2}L^{1-d/2}}\;\,,
\end{equation}
which naturally is independent of temperature.
On a more formal level, the same relation can be obtained as a variational
estimate from above. If one assumes, for example, that
\begin{equation}                                \label{f-trial+}
f(t,{\bf x})=\frac{Ft}{L}
\exp\left(-\frac{{\bf x}^2}{2\Delta_+^2}\right)
\end{equation}
and substitutes Eq. (\ref{f-trial+}) into Eq. (\ref{S(f)0}), then
for $0<d<4$ the result of this substitution $S_{\rm var}(\Delta_+)$ (which
for $F\gg T$ is insensitive to the presence of the viscous term in the
integrand) has a minimum with respect to the variational parameter $\Delta_+$.
This minimum is situated at $\Delta_+(F)$ satisfying relation (\ref{Delta(F)+}),
whereas the value of $S_{\rm var}[\Delta_+(F)]$
satisfies relation (\ref{S(F)+}).

The important difference between the far-left and far-right tails is that
in the far-right tail, the width of the region where the fluctuation of a
random potential is localized grows with the increase in $|F|$.
In such a situation one can expect that the shape of the optimal
fluctuation in the most distant part of the tail at finite $\xi$
will be the same as for a $\delta$-correlated potential.
This requires the fulfillment of the condition $\Delta_+\gg \xi$,
that is, $F\gg J\xi^2/L\,.$
Therefore, for a given $\xi$ and sufficiently large $L$
the region of the applicability of Eq. (\ref{S(F)+}) will be extended
to the whole  non-universal part of the right tail.

Although the minimum of $S_{\rm var}(\Delta_+)$ with respect to $\Delta_+$
exists for any $d$ in the interval $0<d<4$, it follows from the form of
Eq. (\ref{S(F)+}) that this equation can be expected to be directly
applicable only at $d<2$, exactly like in the case of the analogous
expression for the far-left tail, Eq. (\ref{S(F)d}).
For $d>2$ Eq. (\ref{S(F)+})
(where $L$ enters as the total time of the development of the fluctuation)
predicts that the action 
can be decreased by making the time of the development of this fluctuation
much smaller than $L$. According to Eq. (\ref{Delta(F)+}) this
will be accompanied by the decrease in the size of the fluctuation. This
suggests that at $d>2$ the optimal fluctuation must have a different
structure, which has to be sensitive to the form of a random potential
correlator at small lengths.

If the first factor in the right-hand side of Eq. (\ref{f-trial+}) is
replaced by $$\frac{F\sinh(t/L_+)}{\sinh(L/L_+)}\,,$$
which allows one to vary not only the characteristic size of a fluctuation
$\Delta_+$ but also the time of its development $L_+$, then for
$U({\bf x})\propto\exp(-{\bf x}^2/2\xi^2)$ and $d>2$ the minimum of
the action is achieved at $\Delta_+\sim\xi$ and $L_+\sim J\xi^2/F$,
which corresponds to
\begin{equation}                                           \label{S(F)+f}
S(F)\sim \frac{\xi^{d-2}F^3}{U_0 J}\;.
\end{equation}
Note that at $d=2$, the estimates given by Eqs. (\ref{S(F)+}) and (\ref{S(F)+f}) coincide with each other.
Naturally,  at the marginal dimension of $d=2$ (where algebraic divergences
are replaced by logarithmic) some logarithmic factors may appear in the
expression for the action.

\section{Modification of tails by the renormalization effects  \label{ut}}

In terms of the Burgers equation parameters (the viscosity $\nu=T/2J$ and
the pumping force intensity \makebox{$D=U_0/2J^2$}),
Eq. (\ref{S(F)d}) can be rewritten as
\begin{equation}                                        \label{S(F)d2}
    S(F)\sim\frac{\nu^d}{D}
    \frac{(-F/J)^{2-d/2}}{L^{1-d/2}}\;.
\end{equation} 
This estimate has been derived at $d<2$ and $\xi=0$ and
is applicable also at $\xi>0$ as soon as $\xi\ll\Delta$.
However, from the nature of the optimal-fluctuation approach it is clear
that the range of the applicability of Eq. (\ref{S(F)d2}) is
restricted also from the other side because in order to disregard
the renormalization of any parameters by the nonlinearity
the soliton has to be sufficiently narrow:
$\Delta\ll x_0$, where $x_0$ is defined by the relation
\begin{equation}                                        \label{x_0}
x_0^{2-d}\sim\frac{\nu^3}{D}
   \sim\frac{T^3}{JU_0}\;.
\end{equation}
At any $d\neq 2$ $x_0$ is the only parameter with the dimension of ${\bf
x}$ which can be constructed from $T$, $J$ and $U_0$. In particular,
in the case of $d<2$ and small $\xi$ we are discussing now, $x_0$ is the
length scale at which the perturbative corrections to $\nu$ and $D$ become
comparable with the bare values of these parameters.

Thus, at $\Delta\gg x_0$ the renormalization effects become important.
In such a regime the probability of a large negative fluctuation of $F$
is determined not by a single fluctuation (and small deviations from it)
but by a relatively wide class of fluctuations,
the summation over which can be taken into account by analyzing an
optimal fluctuation in a system with renormalized parameters.
Since in all the cases we consider the optimal fluctuations are
quasi-stationary (see below) and well localized at a particular length
scale, this can be done by replacing all parameters in Eq. (\ref{S(F)d2})
by their effective values at the corresponding length scale
and zero frequency. \cite{KK}
However, it is well known that only $\nu$ and $D$ are subject to
renormalization, whereas the amplitude of the nonlinear term in the KPZ
equation (\ref{KPZ}) (and, therefore, the coefficient $J$) cannot
be renormalized as a consequence of the Galilean invariance. \cite{MHKZ}

From the continuity it is clear that when the instanton is not too narrow,
the approach relying on using Eq. (\ref{S(F)d2}) with renormalized
parameters can be also expected to work even at $d\geq 2$ [where Eq.
(\ref{S(F)d2}) has no region of the direct applicability]
as soon as the parameters of the system correspond to the same phase
as at $d<2$ [namely, the strong-coupling phase in which the fluctuations
of $f(t,{\bf x})$ in a stationary situation are divergent,
see Eq. (\ref{ff'}) below]. At $d>2$ this requires to have
$x_0/\xi>\kappa(d)$, that is, the temperature $T$ should be lower then some
critical value $T_c(d)$, \cite{IS} which tends to infinity when
$d\rightarrow 2+0$. In the weak-coupling phase, that is at $T>T_c(d)$,
typical fluctuations of $f(t,{\bf x})$ in the stationary situation can be
described by neglecting the non-linear term in the KPZ equation (\ref{KPZ}).
However, the form of the most distant parts of the tails of $P_L(F)$ is
insensitive to the relation between $T$ and $T_c(d)$ and in both phases has
to be given by Eqs. (\ref{S(F)xi}) and (\ref{S(F)+f}).

To describe how the renormalization effects change the shape of the tails
of $P_L(F)$ in the regime when they are important (which corresponds
to the universal parts of the tails in the strong-coupling phase),
we first have to review some known properties of the stationary solution
of the KPZ model in the strong-coupling regime.

\subsection{Stationary solution of the KPZ model
            \label{ss}}

In a stationary situation the divergence of fluctuations
in the strong-coupling phase of a KPZ system is algebraic.
Their behavior at large scales in space-time can be
described by two fundamental exponents, \cite{MHKZ,HH}
\begin{equation}                                      \label{ff'}
\langle[f(t,{\bf x})-f(t',{\bf x}')]^2\rangle
\propto{|{\bf x}-{\bf x}'|^{2\chi}}
g\left(\frac{|t-t'|}{|{\bf x}-{\bf x}'|^z}\right)\,.
\end{equation}
Here $\chi\equiv\chi(d)$ is the roughening exponent characterizing the
equal-time interface fluctuations, $z\equiv z(d)$ is the dynamic exponent
describing the scaling of the relaxation time with the length-scale,
whereas the function $g(\alpha)$ has a finite limit at $\alpha\rightarrow
0$ and diverges as $\alpha^{2\chi/z}$ when $\alpha\rightarrow\infty$. It
is well known \cite{MHKZ} that the existence of the Galilean invariance
imposes
\begin{equation}                                      \label{z+chi}
z+\chi=2\,.
\end{equation}
At $d=1$ the value of the exponent $\chi=1/2$ is known exactly because
the equal-time correlator of $f(t,{\bf x)}$ in a system with
$\delta$-functional correlations of a random potential has to be exactly
the same as in the absence of the non-linearity. \cite{HHF}
This property is a consequence of the fluctuation-dissipation
theorem, \cite{DH} which is obeyed by Eq. (\ref{KPZ}) only at $d=1$.
At $d\neq 1$ the values of the exponents $z$ and $\chi$ are known only
from approximate or numerical calculations.
In terms of the directed polymer problem the dependence (\ref{ff'})
corresponds to
\begin{equation}                                         \label{x-x'}
\langle[{\bf x}(t)-{\bf x}(t')]^2\rangle\propto (t-t')^{2/z}\,,
\end{equation}
which shows that $\zeta=1/z$  plays the role of the roughening exponent
for the transverse displacements inside an infinite polymer and therefore
cannot be smaller than 1/2, \cite{comm-zeta} from where $z\leq 2$.

A natural way to describe the effective renormalization of  $\nu$ and $D$
by the nonlinearity consist in introducing  \cite{HF}
a generalized viscosity $\nu(\omega,{\bf q})$ and a generalized pumping
intensity $D(\omega,{\bf q})$ defined by the relations
\begin{eqnarray}
                                                       \label{G}
G(\omega,{\bf q}) & = & [-i\omega+\nu(\omega,{\bf q})q^2]^{-1}\,,
\\ C(\omega,{\bf q}) & = & 2J^2|G(\omega,{\bf q})|^2D(\omega,{\bf q})\,,
                                                        \label{C}
\end{eqnarray}
where $G(\omega,{\bf q})$ and $C(\omega,{\bf q})$ are, respectively,
the Fourier transforms of the response function and of the two-point
correlation function of $f(t,{\bf x})$.
The form of Eqs. (\ref{G}) and (\ref{C}) corresponds to the replacement
of the considered non-linear system by a linear system with the same
form of $G(\omega,{\bf q})$ and $C(\omega,{\bf q})$.

The compatibility with the behavior described by Eq. (\ref{ff'})
requires then that at small enough $q$,
\[
\lim_{\omega\rightarrow 0}\nu(\omega,{\bf q})\propto q^{-(2-z)}\,,~~~
\lim_{\omega\rightarrow 0}  D(\omega,{\bf q})\propto q^{-(d+2\chi-z)}\,.
\]
This suggests that the behavior of low-frequency fluctuations
with typical or smaller amplitude can be
qualitatively described by using an effective viscosity
$\nu_{\rm eff}(R)$ and an effective pumping intensity $D_{\rm eff}(R)$
which algebraically depend on a length scale $R$,
\begin{equation}                                       \label{nu_eff}
\nu_{\rm eff}(R)\sim\nu\left(\frac{R}{a_\nu}\right)^{2-z}\hspace*{-4mm},~~~
D_{\rm eff}(R)\sim D\left(\frac{R}{a_D}\right)^{4+d-3z}\hspace*{-8mm},
\end{equation}
where in accordance with Eq. (\ref{z+chi}) we have replaced $\chi$ by $2-z$.
As a convenient way of describing the amplitudes of $\nu_{\rm eff}(R)$
and $D_{\rm eff}(R)$, we have introduced in Eq. (\ref{nu_eff})
two new length scales, $a_\nu$ and $a_D$.
For $d=1$ and $\xi\lesssim x_0$ both $a_\nu$ and $a_D$ can be expected
to be of the order of $x_0$, because in such a situation
$x_0$ is the only relevant length in the problem.
However for $\xi\gg x_0$ and/or $d>1$ these two length-scales do not have
to be of the same order.
Since both $\nu$ and $D$ increase under the renormalization,
Eqs. (\ref{nu_eff}) can be expected to be applicable only for
\makebox{$R\gg a_\nu,a_D$}.

In scaling regime, when $\nu_{\rm eff}(R)$ and $D_{\rm eff}(R)$
behave themselves in accordance with Eqs. (\ref{nu_eff}), both these
quantities have no direct relation to their bare values, $\nu$ and $D$.
Their origin can be traced to the effect of fluctuations with
shorter wave lengths than the given length scale $R$. In particular, it
follows from the structure or the KPZ equation (\ref{KPZ}) that at the
length-scale $R$ the role of the effective random potential is played by
the deviation of $-(J/2)\langle{\bf u}^2\rangle_R$ from its average value,
$-(J/2)\overline{\langle{\bf u}^2\rangle_R}$, where $\langle\ldots\rangle_R$
denotes spatial averaging over a region with a linear size of the order of
$R$. From this the value of $D_{\rm eff}(R)$ can be estimated as
\[
D_{\rm eff}(R)  \sim  \int_{|{\bf r}|<R}d{\bf r}\int_{-\infty}^{+\infty}
d\tau\left[\overline{u^a(t,{\bf x})u^b(t+\tau,{\bf x+r})}\right]^2
\]
\begin{equation}                                             \label{D_eff-2}
\sim  \frac{R^{2+d} u_{\rm typ}^4(R)}{\nu_{\rm eff}(R)}\;.\hspace*{32mm}
\end{equation}
In Eq. (\ref{D_eff-2}) we have assumed that the integration over $d\tau$
can be replaced by the multiplication by the factor $\sim\tau(r)$, where
$\tau(r)\sim r^2/\nu_{\rm eff}(r)$ is the characteristic relaxation time
which can be associated with the length scale $r$, whereas the result of
the integration over $d{\bf r}$ has been estimated assuming that as
a consequence of the universality for any length scale  there exist
only one characteristic velocity scale which can be associated with this
length scale (in other terms, there is no anomalous scaling).
We have chosen as such a velocity scale the typical velocity,
$u_{\rm typ}(R)$, defined by the relation
\begin{equation}                                           \label{u_typ}
u_{\rm typ}^2(R)\equiv\overline{\langle{\bf u}\rangle_R^2}
\sim \frac{D_{\rm eff}(R)}{\nu_{\rm eff}(R)R^d}\;.
\end{equation}

Substitution of Eq. (\ref{u_typ}) into Eq. (\ref{D_eff-2}) then gives the
relation
\begin{equation}                                           \label{nu^3/D}
\frac{\nu_{\rm eff}^3(R)}{D_{\rm eff}(R)}\sim R^{2-d}\,,
\end{equation}
whose structure is analogous to that of Eq. (\ref{x_0}). The consistency
between Eqs. (\ref{nu^3/D}) and (\ref{z+chi}) confirms the correctness
of assumptions which have been used for the derivation of Eq. (\ref{nu^3/D}).
In terms of the length scales $a_\nu$ and $a_D$ introduced above,
see Eqs. (\ref{nu_eff}), relation (\ref{nu^3/D}) can be rewritten as
\begin{equation}                                        \label{a_nu(a_D)}
a_\nu\sim\left(\frac{x_0}{a_D}\right)^\frac{2-d}{3(2-z)}a_D\,,
\end{equation}
which for $d=1$ (when $z=3/2$) is reduced to
\begin{equation}                                        \label{a_nu(a_D)-2}
    a_\nu\sim(x_0^2a_D)^{1/3}\,.
\end{equation}

When the dynamics of fluctuations at $R\sim\xi$ is dominated by wave
breaking, the value of $u_{\rm typ}(\xi)$ can be estimated
as a characteristic velocity
\makebox{$u_\xi\sim(D\tau_{\xi}/\xi^3)^{1/2}$},
which is created by a random force with characteristic length-scale $\xi$
during the time $\tau_\xi\sim\xi/u_\xi$ required for breaking
of such a fluctuation, which gives
\makebox{$u_\xi\sim(D/\xi^{1+d})^{1/3}\,.$}
A comparison of this estimate with Eqs. (\ref{D_eff-2}) and (\ref{u_typ})
suggests that in such a regime $D_{\rm eff}(\xi)\sim D$, that is, $a_D\sim\xi$.
At $d<2$ we expect this conclusion to be applicable when $\xi\gtrsim x_0$,
whereas at $d>2$ - in the whole region of the existence of the
strong-coupling phase.

It follows from the definition of $u_{\rm typ}(R)$
that with the increase in $R$ the value of $u_{\rm typ}(R)$ has to decrease.
A comparison of Eq. (\ref{u_typ}) with Eqs. (\ref{nu_eff}) allows one then
to conclude that $z$ has to be larger than 1.

\subsection{Universal part of the left tail}  \label{ul}

After replacing in Eq. (\ref{Delta(F)}) $T/2J\equiv\nu$ by
$\nu_{\rm eff}(\Delta)$, one obtains a relation which allows one to find
that in the regime when the renormalization effects are important
the estimate for the soliton width $\Delta$ acquires a form
\begin{equation}                                       \label{Delta(F)d}
{\Delta}\equiv\Delta(F)\sim {a_\nu}
\left(\frac{L\nu^2J}{-Fa_\nu^2}\right)^{\frac{1}{2(z-1)}}\,.
\end{equation}
A substantial change of $\Delta$ in comparison with what is given by Eq.
(\ref{Delta(F)}) means that in the regime we consider now the probability
of a large negative fluctuation of $F$ is determined not by a narrow
vicinity of the fluctuation which minimizes the original action (like it
happens in the more distant part of a tail), but by a wide vicinity of an
essentially different fluctuation whose dominance is ensured by a factor
related to the integration over its vicinity.
In the framework of a renormalization group approach this factor is
effectively taken into account when one is replacing different parameters
by their renormalized values.

An estimate for the action can be then obtained by making
in Eq. (\ref{S(F)d2}) a replacement
\begin{equation}                                    \label{nu_eff(Delta)}
\nu\rightarrow\nu_{\rm eff}(\Delta)\,,~~~
D\rightarrow D_{\rm eff}(\Delta)
\end{equation}
with $\Delta$ given by relation (\ref{Delta(F)d}).
With the help of Eq. (\ref{nu^3/D})
the result of this substitution can be reduced to the form
\begin{equation}                                          \label{S(F)u}
S(F) \sim \left(\frac{-F}{F_*}\right)^{\eta}\,,
\end{equation}
with exponent
\begin{equation}                                            \label{eta_-}
\eta=\eta_- 
\equiv\frac{z}{2(z-1)}\;
\end{equation}
which depends on $d$ only through the dynamic exponent $z\equiv z(d)$
but not explicitly.
Here
\begin{equation}                                            \label{F*d}
F_{*}\sim {J\nu}
\left(\frac{\nu L}{a_\nu^2}\right)^\omega 
\end{equation}
plays the role of a characteristic free-energy scale whose dependence
on $L$ is described by the exponent
\begin{equation}                                            \label{omega}
\omega=1-\frac{1}{\eta_-}=\frac{2}{z}-1\;.
\end{equation}

The universality hypothesis for the directed polymer problem \cite{IV}
(or, more generally, for the collective pinning problem \cite{Blatter})
suggests that $F_*$ has to be of the same order as a characteristic
elastic energy $E_{\rm el}\sim J(\delta x)^2/L$, where the dependence of
the characteristic transversal displacement between the two ends of a
polymer, \makebox{$\delta x\equiv |{\bf x}(t=L)-{\bf x}(t=0)|\propto
L^\zeta$}, on its total length $L$ is described by the roughening exponent
$\zeta$ so that
\begin{equation}                                       \label{omega(zeta)}
\omega=2\zeta-1\,.
\end{equation}
A comparison of Eq. (\ref{omega(zeta)}) with Eq. (\ref{omega}) demonstrates
that the fluctuations of $\delta x$ are described by the same roughening
exponent $\zeta=1/z$ as fluctuations inside an infinite polymer, see Eq.
(\ref{x-x'}), in full agreement with what one expects from the
universality. This consistency can be considered as an additional
confirmation of the validity of the set of assumptions which have been
used for obtaining Eq. (\ref{S(F)u}).

Note that the list of these assumptions includes the conjecture that the
system evolves sufficiently slow, so that at relevant length scales it can
be considered as already equilibrated, which is a necessary condition for
using Eqs. (\ref{nu_eff}). For this the total evolution time $L$ has to be
much larger then the characteristic relaxation time
\makebox{$\tau(\Delta)\sim\Delta^2/\nu_{\rm eff}(\Delta)$} which can be
associated with the length scale $\Delta$. \cite{NT} Since in terms of
$\Delta(F)$ and $L$ relation (\ref{S(F)u}) can be rewritten as
\begin{equation}                                        \label{S(tau)}
S(F)\sim \frac{\nu_{\rm eff}(\Delta)}{\Delta^2}L
\sim\frac{L}{\tau(\Delta)}\;,
\end{equation}
the constraint $L\gg\tau(\Delta)$ is equivalent to
$S(F)\gg 1$ and, accordingly, is automatically fulfilled
as soon as one is dealing with the tail.

It is also important that the effective viscosity and effective pumping
intensity given by Eqs. (\ref{nu_eff}) and following from the form
of the correlation function (\ref{ff'}) can be used for
the description only of typical (or more weak) fluctuations.
The comparison of the characteristic velocity of the flow
created around the instanton,
\begin{equation}                                           \label{u_F}
u_F\sim \left(\frac{|F|}{JL}\right)^{1/2}\hspace*{-1mm},
\end{equation}
with $u_{\rm typ}(\Delta)$, the typical velocity of equilibrium
fluctuations at the length scale $\Delta$ [see Eq. (\ref{u_typ})],
demonstrates that in the considered case both quantities are of the same
order, and therefore, the approach based on using Eqs. (\ref{nu_eff}) with
$R\sim\Delta$ is indeed justified. This allows us to conclude that our
instanton is created by fluctuations of the effective random potential
whose amplitude is typical for their length scale. In such a situation the
only reason why the probability of the instanton is small is that the
signs of these typical fluctuations have to be same in all
$L/\tau(\Delta)$ independent time intervals of the length $\tau(\Delta)$.
This provides a qualitative explanation why the expression for the action
can be reduced to a very simple form $S\sim L/\tau(\Delta)$.

At large values of $-F$ the range of the applicability of Eq.
(\ref{S(F)u}) is restricted by the constraint $\Delta(F)\gg a_\nu,a_D$,
which is required for making replacement (\ref{nu_eff(Delta)}). In
particular, when $d<2$ and $\xi\ll x_0$ (so that \makebox{$a_\nu\sim
a_D\sim x_0$}) one can expect that at $\Delta(F)\sim x_0$, that is, at
\begin{equation}                                          \label{Fc}
-F\sim F_c\sim \frac{T^2L}{Jx_0^2}\,,
\end{equation}
a crossover takes place from dependence (\ref{S(F)u})
to dependence (\ref{S(F)d}).

{On the other hand, in situations when $\xi$ (or $d$) is too
large for dependence (\ref{S(F)d}) to have any range of applicability,
one could expect to have a direct crossover between dependences
(\ref{S(F)xi}) and (\ref{S(F)u}). However, the range of
the applicability of Eq. (\ref{S(F)xi}) describing the far-left tail
corresponds to $\Delta(F)\ll \mbox{min}[a_\xi,a_D]$
and of Eq. (\ref{S(F)u}) describing the universal regime
to $\Delta(F)\gg \mbox{max}[a_\xi,a_D]$.
Since we expect that in a general situation the two length scales, $a_\xi$
and $a_D$, are essentially different, we have to admit the existence
in such a case of an intermediate region in $F$, where the form of
the left tail of $P_L(F)$ cannot be established without
further investigation.}

As it has been already mentioned above, at $d=1$ the value of the exponent
$z$ is known exactly. Substitution of $z=3/2$ and $a_\nu\sim x_0$ into Eq.
(\ref{S(F)u}) and Eq. (\ref{eta_-}) then
reproduces an estimate for $S(F)$ which up to unknown numerical factor
coincides with Eq. (\ref{S(F)}) for $S(F)$ in the far-left tail.
This shows that for $d=1$ and $\xi\ll x_0$ the dependence of $S$
on all parameters in the universal part of the left tail
is exactly the same as in its non-universal part at $F_c\ll -F\ll F_\xi$.
In this particular case at $-F\sim F_c$ only a numerical coefficient
in the dependence (\ref{S(F)u}) can experience a crossover.

For $d=1$ and $\xi\gg x_0$,  substitution of Eq. (\ref{a_nu(a_D)-2})
with $a_D\sim\xi$ 
into Eq. (\ref{F*d}) reproduces an estimate for $F_*(L)$ which has been
obtained by Nattermann and Renz \cite{NR} from scaling arguments
complemented by the assumption that at low enough temperatures $F_*(L)$
{has} to be temperature independent, and follows also from the
replica-symmetry-breaking analysis of Ref. \onlinecite{KD}.

For $d\neq 1$ the value of the exponent $\eta$ in the universal part of
the left tail, $\eta_-=z/[2(z-1)]$, does not coincide with its value in
the far-left tail, where it is given by $2-d/2$ [see Eq. (\ref{S(F)d})].
Note that expression (\ref{S(F)u}) decreases with the increase in $L$
as long as $\eta_{-}>1$, that is, $1<z<2$. This means that in the universal
part of the left tail the condition which is necessary for the possibility
of having a macroscopic optimal fluctuation (whose size is much larger
then $\xi$) is changed from $d<2$ to \makebox{$1<z<2$}. On the other hand,
when the renormalization effects are taken into account, the condition
$0<d<4$ required for having a minimum of $S$ with respect to $\Delta$ (see
Sec. \ref{fl-d}) is replaced by $4/3<z<2$. Thus, the range of the
applicability of Eq. (\ref{S(F)u}) is not restricted to $0<d<2$ (as in the
case of the analogous expression for the far-left tail) but extends itself
to the whole region of parameters {where the strong-coupling phase
does exist and $z>4/3$ (the condition $z<2$ always has to be fulfilled,
see Sec. \ref{ss}). Note that for $d=1$ the value of
$\zeta\equiv 1/z$ is equal to 2/3 and according to numerical simulations
goes down with the increase in $d$. \cite{HHZ} This means that
the restriction $z>4/3$ is fulfilled for any physical dimension.}

\subsection{Universal part of the right tail}                \label{ur}

One could expect the approach based on the application
of the replacements (\ref{nu_eff(Delta)}) to be applicable also for the
description of the universal part of the right tail.
However, it turns out that in this case the situation is more complex.
This can be understood by comparing the size of the optimal fluctuation
$\Delta_+(F)$,
given by Eq. (\ref{Delta(F)+}), with the length scale $R_*(F)$ at which
the typical velocity of equilibrium fluctuations $u_{\rm typ}(R_*)$,
given by Eq. (\ref{u_typ}), becomes comparable with
\begin{equation}                                             \label{u_F+}
u_F\sim \frac{F}{J\Delta_+}\,
\sim\left(\frac{F}{JL}\right)^{1/2}\,,
\end{equation}
the characteristic velocity
inside the optimal fluctuation with the size $\Delta_+(F)$.
In Eq. (\ref{u_F+}) we have used the estimate for $\Delta_+(F)$
given by Eq. (\ref{Delta(F)+}), which
has led to exactly the same estimate for $u_F$ in terms of $|F|$
as in the left tail [see Eq. (\ref{u_F})].
This means that in both tails $R_*(F)$ has to be of the same order.
On the other hand, in Sec. \ref{ul} we have established that in the 
left tail the relation $u_F\sim u_{\rm typ}(R_*)$ holds precisely when
$R_*\sim \Delta(F)$. This allows one to conclude  that in the right tail,
\begin{equation}                                             \label{}
R_*(F)\sim \Delta(-F)\,,
\end{equation}
where $\Delta(F)$ is the instanton width in the left tail given by
Eq. (\ref{Delta(F)d}).

Accordingly, for the creation of the optimal fluctuation whose
size $\Delta_+(F)$ is much larger than $R_*(F)$
(as it is required in the case of the right tail),
the fluctuations of the effective random potential with length-scale
$\Delta_+(F)$ should have amplitudes much larger then typical.
Naturally, the probability of such fluctuations is strongly suppressed and
cannot be estimated by using Eqs. (\ref{nu_eff(Delta)}).

The most effective way of formation of a fluctuation whose amplitude $u_F$
substantially exceeds the typical velocity of fluctuations at the
corresponding length-scale consists in formation of a set of fluctuations
with smaller length scales, such that for them the amplitudes
of the order of $u_F$ are typical.
This means that the length-scales of these fluctuations should be
of the order of $R_*(F)$, and, accordingly, the estimate for the action
should include an additional factor $(\Delta_+/R_*)^d$ which takes into
account the need for the spatial coherence of these fluctuations.
This leads to
\begin{equation}                                      \label{S(F)+u}
S(F)\sim \frac{L}{\tau(R_*)}\left(\frac{\Delta_+}{R_*}\right)^d
\sim \left(\frac{F}{F_*}\right)^{\eta_+}\,,
\end{equation}
where $F_*$ is the same characteristic free-energy scale as in
the universal part of the left tail [see Eq. (\ref{F*d})], whereas
exponent $\eta_+$ is given by
\begin{equation}                                   \label{eta+u}
\eta_+ = \frac{(1+d)z}{2(z-1)}\;\,.
\end{equation}
In terms of the renormalization approach exactly the same result
is obtained if the renormalization is stopped not at the scale
$\Delta_+(F)$, corresponding to the total size of the optimal fluctuation,
but at a smaller scale $R_*$ (at which the fluctuations stop to be strong
enough for inducing the renormalization), that is,
by using Eq. (\ref{S(F)+}) with the replacement
\begin{equation}                                      \label{D_eff}
D\rightarrow D_{\rm eff}(R_*)\sim D\left(\frac{R_*}{a_D}\right)^{4+d-3z},
\end{equation}
where $R_*\sim \Delta(-F)$.
Since the value of $D_{\rm eff}(R_*)$ does not depend on $\Delta_+$,
the condition for the existence of a minimum of $S$ with respect to
$\Delta_+$ remains the same as has been found when deriving
Eq. (\ref{S(F)+}), $0<d<4$.
On the other hand, in the universal part of the right tail the condition
required for the possibility of having a macroscopic optimal fluctuation
(whose size is much larger then $\xi$) is changed from $d<2$ to $1<z<2$,
which in the strong-coupling phase anyway has to be fulfilled [see
Sec. \ref{ss}].
Therefore, the range of the applicability of Eq. (\ref{S(F)+u}) is
restricted from above not by $d=2$ (as in the case of the analogous
expression for the far-right tail) but by $d=4$.

Note that in contrast to exponent $\eta_-$ given by Eq. (\ref{eta_-}),
exponent $\eta_+$ depends both on $z$ and $d$.
However, the ratio of these two exponents does not depend on $z$,
\begin{equation}                                         \label{}
\frac{\eta_+}{\eta_-}=1+d\;,
\end{equation}
and therefore is known exactly.
The fact that in the regime where the renormalization effects are important
both tails of the free energy distribution function incorporate
the same characteristic free-energy scale $F_*$ confirms that
this regime corresponds to studying the universal form of this
distribution function.

A comparison of Eq. (\ref{Delta(F)+}) with Eq. (\ref{Delta(F)d}) allows one
to verify that the condition \makebox{$\Delta_+(F)\gg R_*(F)$,} on which
we have relied when deriving Eq. (\ref{S(F)+u}), is equivalent to
$S(F)\gg 1$, and therefore is always satisfied as soon as we are 
dealing with the tail.
Another condition whose fulfillment is required to justify
replacement (\ref{D_eff}) is related to the quasistationarity of the
problem. Namely, the total evolution time $L$ has to be much larger than
the characteristic relaxation time $\tau(R_*)\sim R_*^2/\nu_{\rm eff}(R_*)$
which can associated with the length scale $R_*(F)$.
For $R_*(F)\sim\Delta(-F)$, this condition is also reduced to $S(F)\gg 1$.

From the side of large $F$ the range of the applicability of Eq.
(\ref{S(F)+u}) is restricted by {the constraint $R_*\gg a_D$, whose
fulfillment is also required for making replacement (\ref{D_eff}). In
particular, at $d<2$ and $\xi\lesssim x_0$ (when $a_D\sim x_0$), the
crossover between dependences (\ref{S(F)+u}) and (\ref{S(F)+}) can be
expected to occur at $F\sim F_c$, where $F_c$ is given by the same
expression [Eq. (\ref{Fc})] as in the left tail. On the other hand, at
$d>2$ the crossover between dependences (\ref{S(F)+u}) and
(\ref{S(F)+f}) has to take place while $R_*(F)$ is still much
larger than $\xi$. In this situation we expect that the two contributions
to $P_L(F)$ [one from the ``macroscopic" instanton, corresponding to
dependence (\ref{S(F)+u}) and the other from the ``microscopic" instanton
corresponding to dependence (\ref{S(F)+f})] can coexist with each
other and the crossover has to occur when they become comparable
with each other.}

\section{Conclusion \label{Concl}}

In the present work we have studied the form of the tails of the
free-energy distribution function $P_L(F)$ in the directed polymer problem
both for a $\delta$-correlated random potential and for the case of
a finite correlation length $\xi$.
In all regimes that we  have investigated the tails have
a stretched-exponential form,
\begin{equation}                                           \label{}
-\ln P_L(\pm F)\sim \left[\frac{F}{F_*(L)}\right]^{\eta_\pm}\,,
\end{equation}
with $F_*(L)\propto L^{\omega_\pm}$ and therefore can be characterized by
the two exponents whose values  depend on the dimensionality
of the space in which the polymer is imbedded. We use letter $d$ to denote
the  transverse dimensionality of this space, that is, the number of
components of the displacement vector ${\bf u}$.

For sufficiently large fluctuations of $F$
the form of the tails of $P_L(F)$ is
determined by the form of the most optimal fluctuation of a random potential
which is sufficient for achieving a given value of $F$.
For a $\delta$-correlated random potential and $d<2$ the minimization
of the action corresponding to such a fluctuation
allows one to show that in the far-left tail
\begin{equation}                                             \label{c-fl}
\eta_-=\frac{4-d}{2}\;,~~~\omega_-=\frac{2-d}{4-d}\;.
\end{equation}
The same values of $\eta_-$ and $\omega_-$ have been obtained by Monthus
and Garel \cite{MG} by constructing a generalization
of the Imry-Ma scaling argument (based on a disorder-dependent
Gaussian variational approach introduced in Ref. \onlinecite{GO}).
However, the approach of Ref. \onlinecite{MG} leaves one in doubt on
what is the range of its applicability (and if such a range exists at all),
whereas the methods used in this work allowed us to establish
that the exponents (\ref{c-fl}) are applicable in the most distant part of
the left tail corresponding to the nonuniversal regime.

At $d\geq 2$ the problem with strictly $\delta$-functional correlations
of a random potential becomes ill-defined,
so it becomes necessary to introduce some regularization.
The natural way of doing it consists in assuming that a random potential
correlations are characterized by a finite correlation radius $\xi$.
In the case of $\xi>0$ one finds that in the most distant part of
the left tail the size of the optimal fluctuation of a random potential
has to be comparable with $\xi$ and the values of the exponents become
superuniversal, that is, not dependent on $d$,
\begin{equation}                                        \label{c-fl-xi}
\eta_-=2\,,~~~\omega_-=1/2\,.
\end{equation}
For $d<2$ and not too large $\xi$ one can expect to have a
crossover from regime (\ref{c-fl-xi}) to regime (\ref{c-fl}).

The application of the optimal-fluctuation approach
to the analysis of the right tail shows that for $d<2$
the most distant part of this tail is described by
\begin{equation}                                        \label{c-fr}
\eta_+=\frac{4+d}{2}\,,~~~\omega_+=\frac{2-d}{4+d}\,.
\end{equation}
In contrast to the case of the left tail, the form of the most distant
part of the right tail is insensitive to whether $\xi$ is zero or finite.
On the other hand, for $d>2$ the size of the optimal fluctuation again
becomes determined by $\xi$, which leads to the change of the exponents to
\begin{equation}                                        \label{}
\eta_+=3\,,~~~\omega_+=0\,.
\end{equation}

Note that the value of $\omega_+$ given by Eq. (\ref{c-fr}) corresponds
to the value of the roughening exponent,
\begin{equation}                                        \label{}
\zeta_{\rm F} 
=\frac{3}{4+d}\;\;,
\end{equation}
which is known as ``Flory exponent'' \cite{Kard-JAP} and follows
from simple scaling arguments of Refs. \onlinecite{Kard-JAP}, as well as
from the Gaussian variational calculation of Mezard and Parisi \cite{MP}
incorporating a hierarchical replica-symmetry breaking.
Our analysis has revealed that this scaling analysis (which insofar has been
assumed to be of little relevance, because it cannot reproduce the exactly
known value of $\zeta=2/3$ at $d=1$) in reality is applicable for
the description of the most distant (non-universal) part of the right
tail of $P_L(F)$.
However, it still remains unclear whether the appearance of the same
exponent in the variational calculation of Ref. \onlinecite{MP} (based on
the {\em maximization} of the variational free energy of a system with
$L=\infty$) is not more than a coincidence.

If the parameters of the system correspond to the strong coupling phase,
the decrease in $|F|$ makes the optimal-fluctuation approach no
longer directly applicable because the size of the optimal fluctuation
becomes too large (or its amplitude becomes too small) to neglect
the renormalization of the parameters of the system by fluctuations.
In such a situation, a consistent inclusion of the renormalization effects
into account allows one to express the exponents in terms of the roughening
exponent $\zeta=1/z$ describing the behavior of displacement
fluctuations inside an infinite polymer [see Eq. (\ref{x-x'})].
For universal parts of left and right tails, one obtains, respectively,
\begin{equation}                                     \label{c-eta-pm}
\eta_-=\frac{1}{2(1-\zeta)}\;,~~~\eta_+=\frac{1+d}{2(1-\zeta)}\;\,.
\end{equation}
Not unexpectedly, one finds that the value of $\omega$ is the same for
both tails and is equal to $2\zeta-1$,
as it could be expected from the universality.
Quite remarkably, the ratio $\eta_+/\eta_-=1+d$ does not depend on $\zeta$.

The value of $\zeta$ is known exactly only at $d=1$, where $\zeta=2/3$.
In this case the values of $\eta_-=3/2$ and $\eta_+=3$  which
follow from Eqs. (\ref{c-eta-pm}) are in perfect agreement with
the exact solution \cite{PS} of the polynuclear growth (PNG) model,
which is accepted to belong to the same universality class.
In terms of the directed polymer problem the PNG model
corresponds to the Poisson distribution of identical pointlike impurities
and a rather peculiar limit of vanishing elasticity, $J=0$, and
zero temperature. \cite{PS,Johan}
For this model the form of the distribution function $P_L(F)$ in the universal
regime, as well as the scaling function $g(\alpha)$ entering Eq. (\ref{ff'}),
is known exactly. \cite{PS,PS-04}
The consistency between our results and that of Ref. \onlinecite{PS}
confirms that the directed polymer problem defined by Eq. (\ref{H}) and
the PNG model indeed belong to the same universality class.

The nonuniversal tails in the PNG model have been analyzed in
Ref. \onlinecite{DZ}. Naturally, in the nonuniversal regime even the models
belonging to the same universality class can have different tails.
The difference is especially evident in the case of what we call
the far-right tail because in the PNG model the energy is by definition
bounded from above and therefore its distribution function has to vanish
for large enough positive fluctuations.
On the other hand, it follows from Ref. \onlinecite{DZ}
that in the PNG model the far-left tail is described by
$S(F)\propto F\ln(-F/L)$ and, thus, also has nothing in common with
the far-left tail of the model considered in this work.

In terms of the exponent $\omega=2\zeta-1$ Eqs. (\ref{c-eta-pm})
can be rewritten as
\begin{equation}                                     \label{c-eta-pm2}
\eta_-=\frac{1}{1-\omega}\;,~~~\eta_+=\frac{1+d}{1-\omega}\;\,.
\end{equation}
Our results demonstrate that in model (\ref{H})
the analogous relations are fulfilled also
in non-universal regimes (where $\omega$ is not obliged to coincide
with $2\zeta-1$ and be the same in both tails)
as soon as the size of the optimal fluctuation
is comparable with the total length of a string.
For the far-left tail this has been known \cite{HHZ,MG-07}
from the Kardar-Zhang replica approach.
Recently both relations (\ref{c-eta-pm2}) have been derived by
Mothus and Garel \cite{MG-07} with the help of a recursive procedure
for the zero-temperature problem on a hierarchical diamond
lattice whose effective dimension is equal to $d$.
These authors have also suggested that the same relations can be
expected to hold on all hypercubic lattices.
Although, in our opinion, the argument accompanying this proposal
does not take into  account some important differences between
hypercubic and hierarchical lattices, the results derived in this
work confirm its correctness.


\section*{ACKNOWLEDGEMENTS}

The authors are grateful to G. Blatter, T. Garel, V.B. Geshkenbein,
A.I. Larkin and V.V. Lebedev for useful discussions.
The work of I.V.K. was supported by RFBR under Grant No. 06-02-17408-a.

\appendix
\section{The replica approach \label{RA}}

The replica approach to the directed polymer problem
is based on calculating the moments $Z_n\equiv \overline{Z^n}$
of the distribution of the partition function $Z\equiv z(L,0)$
and allows one to find the far-left tail of $P_L(F)$
without relying on the analytical continuation of $n$ to 0.
Kardar \cite{Kardar} was the first to notice that for any integer $n>1$
(and large enough polymer length
$L$) $Z_n$ with an exponential accuracy
can be approximated as
\begin{equation}                                         \label{Zn0}
Z_n\approx\exp[-E_0(n)L/T]\,,
\end{equation}
where $E_0(n)$ is the ground state energy of
the quantum-mechanical Hamiltonian,
\begin{equation}                                         \label{Hqn}
    \hat{H_n}=-\frac{T^2}{2J}\sum_{a=1}^n\nabla_a^2-\frac{1}{2T}
    \sum_{a=1}^n\sum_{b=1}^n U({\bf x}_a-{\bf x}_b),
\end{equation}
describing $n$ bosons whose mass is equal to $J$ (with $T$ playing the role
of $\hbar$) and interaction to $-U({\bf x})/T$.
In a $(1+1)\,$-dimensional system with a $\delta$-correlated random
potential, \makebox{$U(x)=U_0\delta(x)$},
the ground-state wave-function 
for the Hamiltonian (\ref{Hqn}) and its energy,
\begin{equation}                                         \label{}
E_0(n)=-\frac{U(0)}{T}n-\frac{JU_0^2}{24T^4}n(n^2-1)\,,
\end{equation}
can be found exactly. \cite{Thacker} This gives
\begin{equation}                                         \label{Zn}
Z_n\propto\exp\left(\frac{JU_0^2}{24T^5}n^3L\right)\,,
\end{equation}
where the linear in $n$ term in $E_0(n)$ has been omitted, because it can
be eliminated by a constant shift of the potential $V(t,{\bf x})$
in Eq. (\ref{H}). Note that the form of Eq. (\ref{Zn}) does not depend on
the initial condition. The choice of the initial condition manifests
itself only in the form of a prefactor which in the first approximation
can be ignored.

Since $Z^n\equiv\exp(-nF/T)$, ${Z_n}$ can also be expressed
in terms of $P_L(F)$,
the distribution function of the free energy $F\equiv f(L,0)=-T\ln Z$:
\begin{equation}                                       \label{expF}
Z_n=\int_{-\infty}^{+\infty}dF\,P_L(F)\,\exp(-nF/T)\,.
\end{equation}
It is clear that the integral in Eq. (\ref{expF}) can be convergent
for any integer $n>1$ only if the left tail of $P_L(F)$ decays
faster than exponentially.
In such a case the integration in Eq. (\ref{expF}) can be performed
with the help of the saddle point approximation.
A comparison of the two expressions for $Z_n$
[Eq. (\ref{Zn}) and Eq. (\ref{expF})] has led Zhang
to conclusion \cite{Zhang} that the algebraic growth of
$\ln Z_n\propto Ln^{1/\omega}$ at large $n$ can be recovered only if one
assumes that when $-F$ is large $S(F)\equiv -\ln[P_L(F)]$ is proportional
to $(-F/L^{\omega})^{\eta}$ with $\eta=1/(1-\omega)$. \cite{HHZ}
Thus, $1/\omega=3$ corresponds to $\eta=3/2$.

It is not hard to check that expression  (\ref{S(F)}) for $S(F)$
derived in Sec. \ref{d=1} allows one to reproduce not only the power of
$n$ but also the full expression (\ref{Zn}), including the coefficient in front
of $n^3$.
For such a form of the tail the integral in Eq. (\ref{expF}) at  positive
$n$ is dominated by the vicinity of the saddle-point situated at
\begin{equation}                                      \label{F0}
F= -F_c(L)n^2\,,~~~F_c(L)\equiv \frac{JU_0^2L}{8T^4}\,,
\end{equation}
where the full expression standing in the exponent,
\begin{equation}                                      \label{tildeS}
\tilde{S}(F)=-S(F)-nF/T\,,
\end{equation}
has a maximum with respect to $F$.
Substitution of Eq. (\ref{F0}) into Eq. (\ref{tildeS}) leads then to Eq.
(\ref{Zn}), whereas calculation of
$\partial^2 \tilde{S}(F)/\partial F^2=1/(2TF_c n)\,$
allows one to verify that the condition for the applicability of the
saddle-point approximation has a form
$(F_c/T)n^3\gg 1$ and is automatically fulfilled for any integer $n>1$ as
soon as $F_c\gg T$, which anyway is required for the applicability of
Eq. (\ref{Zn0}). 

Since the explicit expression for $Z_n$ given by Eq. (\ref{Zn}) can be
derived only at integer \makebox{$n>1$},
the region of the applicability of the replica approach is restricted to
\makebox{$-F\gg F_c(L)$}, that is, coincides with the region where the left
tail can be found with the help of the optimal-fluctuation approach (see
Sec. \ref{d=1}) without taking into account the renormalization effects.
However, the analysis of Sec. \ref{ul} has demonstrated that in
the $(1+1)\,$-dimensional systems with $\delta$-functional correlations
the form of the left tail in the region $F_*\ll -F\ll F_c$, where the
renormalization effects have to be taken into account, remains
qualitatively the same as for $-F\gg F_c$. In terms of the replica
approach this means that the analytical continuation of the partition
function of model (\ref{Hqn}) even at $n<1$ behaves itself
as if it was dominated by the contribution from the same state as at $n>1$,
although at $n<1$ this state no longer has the lowest energy. \cite{DIGKB}
The reasons for that still remain to be elucidated.

For $\xi>0$ the form of the ground state of the Hamiltonian (\ref{Hqn})
cannot be found exactly even when $d=1$. However, for $\xi\gg x_0$ the
value of $E_0(n)$ in a $(1+1)$-dimensional system with $\xi>0$ can be
found rather accurately for any integer $n\geq 1$, because in this regime
the typical distance between bosons is much smaller then the radius of
their interaction \cite{KD} and thus the main contribution to $E_0(n)$ is
given simply by $-U(0)n^2/2T$. A comparison of $Z_n\sim\exp[U(0)Ln^2/2T^2]$
with Eq. (\ref{expF}) then immediately leads to the conclusion that the
form of the far-left tail must be described by $S=F^2/2U(0)L$, as it has
been already derived in Sec. \ref{frc} with the help of
the optimal-fluctuation approach. From the origin of this result
it is clear that it has to be applicable for the description
of the most distant part of the left tail at finite $\xi$ for any $d$.

The two contradicting attempts of generalizing the replica approach to
$d>1$ in the regime when the main contribution to $E_0(n)$ is determined
by the full form of $U({\bf x})$ (and therefore cannot be found without
introducing some additional approximations) have been undertaken by Zhang
\cite{Zhang90} and Kolomeisky. \cite{Kolom}

\section{Fixed initial condition \label{FIC}}

If at $t=0$ the polymer is fastened at $x=0$, the initial condition
for the partition function $z(t,x)$ has to be written as
\begin{equation}                                    \label{f-ini-z}
z(0,x)=x_d\delta(x)\,,
\end{equation}
where a dimensional factor $x_d$ (with the same dimensionality as $x$)
has to be inserted to make $z(t,x)$ a dimensionless quantity.
This factor has no physical meaning and must drop out from all
physical quantities.

In the absence of pumping the initial condition (\ref{f-ini-z}) leads to
\begin{equation}                                 \label{}
z(t,x)
=\frac{x_d}{(4\pi\nu t)^{1/2}}\exp\left[-\frac{x^2}{4\nu t}\right]
\equiv z^{(0)}(t,x)
\end{equation}
and
\begin{equation}                                 \label{f0}
f(t,x)=
\frac{Jx^2}{2t}+\frac{T}{2}\ln\frac{4\pi\nu t}{x_d^2}
\equiv f^{(0)}(t,x)\,,
\end{equation}
which in terms of $u\equiv\nabla_x f(t,x)/J$
corresponds to
\begin{equation}                                \label{u0}
    u(t,x)=x/t\equiv u^{(0)}(t,x)\,.
\end{equation}
Accordingly, the initial condition for $f(t,x)$ can be formulated as
\begin{equation}                             \label{f-ini}
\lim_{t\rightarrow 0}
\left[f(t,x)-f^{(0)}(t,x)\right]=0\,.
\end{equation}
In such a situation it seems to be convenient to count off the free energy
from its value in the absence of pumping introducing
\begin{equation}                             \label{}
\tilde{f}(t,x)=f(t,x)-f^{(0)}(t,x)\,,
\end{equation}
(i) because $\tilde{f}(x,t)$ in contrast to $f(t,x)$
does not depend on $x_d$ and (ii) because this allows
to rewrite the initial condition (\ref{f-ini}) as
\begin{equation}                             \label{}
\tilde{f}(0,x)=0\,.
\end{equation}
Accordingly, in the case of fixed initial condition $F$ should be
redefined as
\begin{equation}                             \label{}
F=\tilde{f}(L,0)=f(L,0)-f^{(0)}(L,0)\,.
\end{equation}
It is clear that the stationary solution (\ref{f(x)})
in no way resembles the initial condition (\ref{f-ini}).
However, like in the case of the free initial condition,
it turns out possible to modify this solution
{\em without increasing the action}
in a way which eliminates the inconsistency
with the initial condition.

This solution has to interpolate between the soliton at small $x$
and the dependence \makebox{$f(t,x)=f^{(0)}(t,x)$}, which has to survive
in the regions that in a given moment are too far from the soliton to feel
its presence.
One can expect that the time of the formation of a solution, $t_1$,
will be of the same order as for free initial condition,  $t_1\sim\nu/v^2$,
because at $t\gg t_1$ and $x\sim\Delta$ the velocity of the flow
(\ref{u0}) produced by fixed initial condition is much smaller then the
velocity $v\sim\nu/\Delta$ produced by the soliton with the width $\Delta$.
Then at $t\gg t_1$ and $\Delta\ll x\ll vt$ $f$ has to be of the form
\begin{equation}                                          \label{fs}
f(t,x)=-\frac{Jv^2}{2}(t-t_1)+Jv|x|
+\frac{T}{2}\ln\frac{4\pi\nu t_1}{x_d^2}
\equiv f^{\rm (s)}(t,x)\;,
\end{equation}
where the constant has been chosen in such a way that the extrapolation
to $t=t_1$ gives $f(t_1,0)=f^{(0)}(t_1,0)$.

A comparison of Eq. (\ref{f0}) with Eq. (\ref{fs})
shows that the crossover between these two dependences has to take place
in the vicinity of $|x|=vt$.
Since we assume that at $|x|\gg\Delta$ the potential is absent, this
crossover has to be described by
\begin{equation}                                          \label{fc}
f(t,x) = f^{\rm (s)}(t,x)-T\ln z_1(t,|x|-vt)\,,
\end{equation}
where the function $z_1(t,x)$
is the exact solution of the diffusion equation $z_t=\nu z_{xx}$
which at $t=t'_1=t_1\exp[(Jv^2/2T)t_1]\sim t_1$
[that is, when $f^{\rm (s)}(t,vt)$ coincides with $f^{(0)}(t,vt)$]
smoothly interpolates between $z_1\approx 1$ at $-x\gg\Delta$
and $z_1\approx\exp[-Jx^2/2Tt'_1]$ at $x\gg \Delta$.
Therefore, the asymptotic behavior of $z_1(t,x)$ at large $t$
can be found by assuming
\begin{equation}                                          \label{z1}
z_1(t_1',x)=\theta(-x)+\theta(x)\exp[-Jx^2/2Tt_1']\,,
\end{equation}
where $\theta(x)=\frac{1}{2}[1+\mbox{sign}(x)]$ is the step-like function.
However, for establishing the relation between $S$ and $F$ the exact form
of $z_1(t,x)$ is irrelevant.
We only have to be sure that at $x\approx -vt$ this quantity is very close
to 1, and this is satisfied as soon as $t\gg t'_1.$

The main difference with the case of free initial condition appears in the
relation between $\Delta$ and $F$.
Subtraction of Eq. (\ref{f0}) from Eq. (\ref{fs}) shows that for fixed
initial condition Eq. (\ref{F(Delta)}) should be replaced by
\begin{equation}                                       \label{F(Delta)f}
    -F=\frac{T^2}{2J\Delta^2}L+\frac{T}{2}\ln\frac{L}{t_1}\,.
\end{equation}
However, for $L\gg t_1$ the second term in the right-hand side
of Eq. (\ref{F(Delta)}) is much smaller then the first one
(which is of the order of $TL/t_1$) and therefore can be neglected.

This allows one to conclude that in the case of fixed initial condition
the main contribution to the action has exactly the same form as for
free initial condition. The derivation above can be easily generalized for
the case of $d>1$, as soon as one can assume that the optimal
fluctuation of a random potential remains almost uniform along $t$.


\begin{thebibliography}{99}
\bibitem{comm-zeta} Thermal fluctuations lead to $\zeta=1/2$ even if
                  a disorder is absent.
\end{thebibliography}
\end{document}